\documentclass[ms]{aastex}
\usepackage{spr-astr-addons}
\usepackage{url}

\RequirePackage{color}

\renewcommand{\vec}[1]{\mbox{\boldmath $#1$}}

\newcommand{\m}{{\ \mathrm m} }
\newcommand{\s}{{\ \mathrm s} }
\newcommand{\e}{{\ \mathrm e} }
\newcommand{\kg}{{\ \mathrm {kg}} }
\newcommand{\km}{{\ \mathrm {km}} }

\newcommand{\emaila}{wilhelm@mps.mpg.de}
\newcommand{\emailb}{bholadwivedi@gmail.com}

\begin{document}

\title{Understanding disk galaxy rotation velocities without dark
matter contribution\,--\,a physical process for MOND?
}
\shorttitle{Radial acceleration of disk galaxies}
\shortauthors{K. Wilhelm, B.N. Dwivedi}

\author{Klaus Wilhelm}
\affil{Max-Planck-Institut f\"ur Son\-nen\-sy\-stem\-for\-schung
(MPS), Justus-von-Liebig-Weg 3, 37077 G\"ottingen, Germany \\ \emaila}

\and

\author{Bhola N. Dwivedi}
\affil{Department of Physics, Indian Institute of Technology
(Banaras Hindu University), Varanasi-221005, India \\ \emailb}

Last updated on \today

\vspace{1cm}

\begin{abstract}
An impact model of gravity designed to emulate Newton's law of gravitation is
applied to the radial acceleration of disk galaxies. Based on this model
\citep{Wiletal}, the rotation velocity curves can be understood without
the need to postulate any dark matter contribution. The increased acceleration
in the plane of the disk is a consequence of multiple interactions of gravitons
(called "quadrupoles" in the original paper) and the subsequent propagation
in this plane and not in three-dimensional space.
The concept provides a physical process that relates the fit parameter
of the acceleration scale defined by \citet{McGetal} to the mean free path
length of gravitons in the disks of galaxies. It may also explain
the modification of the gravitational interaction at low acceleration levels
in MOND \citep{Mil83,Mil94,Mil15,Mil16}. Three examples are discussed
in some detail: The spiral galaxies NGC\,7814, NGC\,6503 and M\,33.
\end{abstract}

\keywords{Disk galaxies, rotation curves, modified gravity}

\section{Introduction} 
\label{s.introd}
Since \citet{Oor32} and \citet{Zwi33} introduced the concept of
\emph{Dark Matter} (DM; \emph{dunkle Materie}), because (1) of discrepancies of
velocity distributions in the Milky Way Galaxy and (2) the large speed
deviations within the Coma galaxy cluster that are in conflict with its
stability, respectively, DM concepts have been applied, in particular, to the
flat rotation curves of disk galaxies, but also to the anomalous deflection of
light \citep[cf., e.g.,][]{Rub83,Rub86,Ell10}.

We mainly consider the rotation dynamics of galaxies and propose
a physical process which explains the observations without the need to
introduce any DM.

\citet{Binetal} states that there is independent
`global' and `local' evidence for galactic DM and asks if the same
mysterious material can be responsible in both cases. We will not address the
global evidence, but question the local one.
As far as the global aspect is concerned, hundreds of articles discuss
\emph{Cold Dark Matter} (CDM) models of cosmology with DM halos
dominating the dynamics of the universe and\,--\,combined with
the $\Lambda$CDM paradigm, where $\Lambda$ is the cosmological
constant\,--\,are successful in
explaining both the large-scale cosmic structures as well
as galaxy formation and evolution over an enormous span
of redshifts \citep[e.g.][]{Navetal,Mooetal,Weietal,Ohetal}.
Nevertheless, important features related to the nature
and origin of DM are still missing and the empirical knowledge of
dark halos remains very sparse \citep{Bur95,TreKoo,Boyetal}.

The interesting question:
`Are the tensions between CDM predictions and observations
on  the  scales  of  galactic  cores  and  satellite  halos  telling  us
something about the fundamental properties of dark matter, or are
they telling us something interesting about the complexities of
galaxy formation?'
asked by \citet{Weietal} leads to
their suggestion that investigating the influence of baryons on the DM halo
profiles could be a direction for future progress.

\citet{McG05a} observed
a fine balance between baryonic and dark mass in spiral galaxies that may
point to new physics for DM or a modification of gravity. \citet{Fraetal}
have also concluded that either the baryons dominate the DM
or the DM is closely coupled with the luminous component. \citet{Sam16}
could describe the dynamics of NGC\,5128 both with a DM halo
and MOND.
\citet{SalTur} have suggested that there is a profound interconnection between the
dark and the stellar components in galaxies.
\citet{McGetal} have presented a correlation between the radial acceleration
and the observed distribution of baryons in 153 rotationally supported
galaxies and have concluded that the DM contribution is fully specified by that of
the baryons \citep[cf.][]{Mil16}. \citet{Sal17} supports this correlation,
but sees no reason that it challenges the DM scenario in galaxies.
A brief review of the literature on
spiral galaxies is required, before we will argue that the DM scenario in
galaxies is not required.

\section{Spiral galaxies} 
\label{s.spirales}
\subsection{Flat rotation curves} 
\label{ss.flat}
From the flat or slightly increasing rotation curves, \citet{Rub86} has
concluded that the DM is clumped about galaxies in a mostly spherical halo
and is more extended than the
luminous matter.
\citet{Binetal}
expected the halo to be flattened in the same way although not
to the same extent as the disk.
Recent observations indicate for the disk light,
a flattening ratio of 7.3 \citep{Kreetal}, and
for seven dwarf galaxies seen edge-on an average stellar disk scale
length to height ratio of $\approx 9$ \citep{Petetal}.
The flat rotation curves seem to require
that the mass interior to $R$, $M(R)$ is proportional to $R$.
The observations\,--\,described, for instance, by \citet{Sal08}\,--\,that
the rotation curves of spiral galaxies do not show a Keplerian fall-off and,
therefore, do not match the gravitational
effects of the stellar plus gaseous matter are supported by many other
studies \citep[cf.][]{CorSal,McG05b,Greetal,Coretal,McGetal,Kam17}.

The observation that the circular velocity
at large radii~$R$ around a finite galaxy becomes independent
of $R$\,--\,resulting in asymptotically flat velocity curves\,--\,led
\citet{Mil83,Mil94} to
propose a modification of the gravitational interaction at low acceleration
levels, called
MOdification of the Newtonian Dynamics (MOND). Recent discussions
and a review indicate that MOND might describe the dynamics of spiral galaxies
without DM \citep[e.g.,][]{Gir00a,FamMcG,Angetal,Kroetal}.

\subsection{Core-cusp problem} 
\label{ss.Core_cusp}
\citet{deB10} has provided a clear exposition of the problem:
`This paper gives an overview of the attempts to determine the
distribution of dark matter in low surface
brightness disk and gas-rich dwarf galaxies, both through observations and
computer simulations.  Observations  seem  to  indicate  an  approximately
constant dark  matter  density  in  the  inner  parts  of  galaxies,
while cosmological computer simulations indicate a steep power-law-like
behaviour.  This difference has become
known as the ``core/cusp problem'', and remains one of the unsolved problems
in small-scale cosmology.'

The core-cusp problem therefore is one of the main subjects of many papers
on dwarf galaxies \citep[e.g.,][]{NavSte,Klyetal4,TreKoo,Genetal,HagWil,
DeGetal,Ohetal,Ebyetal,Kam17}.
Even warm DM does not do better than cold DM in solving the
small-scale inconsistencies \citep{Sch14}.
\citet{Lopetal} find that the Navarro-Frenk-White \citep[NFW,][]{Navetal,NFW}
DM profile provides a better fit to the rotation curve data than
the cored Burkert one \citep{Bur95}.
\citet{BotPes} conclude that
the central regions of galaxies definitively have cores
and not cusps. The core radii of the resultant halos are typically one
to a few kiloparsec, apparently independent of galaxy mass.

\subsection{Satellite galaxies} 
\label{ss.S_gal}
Further discrepancies between observations and numerical simulations
of $\Lambda$CDM and CDM models of cosmology are (1) the ``missing satellites
problem'', i.e., the number of satellite galaxies of the Milky Way and other
large galaxies is far fewer than the number predicted in simulations,
and (2) that the observed number of dwarf galaxies
as a function of rotation velocity is smaller than predicted by the standard
model\,--\,the so-called ``too big to fail'' problem \citep{
Klyetal,Moo01,Kra10,Boyetal,Toletal,Kroetal,StrWec,BotPes,
Ebyetal,Ohetal,Sch17}. Moreover, the dwarf galaxies are not isotropically
distributed around the Andromeda Galaxy and the Milky Way, but show `a highly
flattened distribution around each of the galaxies' \citep{Henetal}.

\subsection{New ideas} 
\label{ss.ideas}
In view of the many discrepancies between the standard cosmological model
and the observational data obtained, in particular, from dwarf
galaxies, many new ideas have been discussed in the literature, and
\citet{Wil01} writes: `We can identify deep questions that seem to call for
ideas outside our present grasp'.

\citet{Weietal} consider as alternatives that the small-scale conflicts could
be evidence of more complex DM physics itself or that baryonic effects can
account for some of the discrepancies; and
a Whitepaper describes the new science opportunities and experimental
possibilities for the exploration of the nature of DM by small projects
\citep{Batetal}.

\section{Baryons and Dark Matter} 
\label{s.baryons}
\citet[][see Equation~9]{McG12} has derived for many galaxies
the Baryonic Tully-Fisher Relation
(BTFR), which implies that the baryonic mass~$M_{\rm bar}$ of a spiral galaxy
is related to the flat rotational velocity~$v_{\rm f}$,
%
\begin{equation}
M_{\rm bar}(v_{\rm f}) = A\,v^4_{\rm f}~,
\label{eqn.Tully_Fisher}
\end{equation}
where $A = (9.4 \pm 1.2) \times 10^{19}\kg\m^{-4}\s^4$
provides a good fit.
Any DM contribution in
rotationally supported galaxies would be specified by the baryon contribution
according to Equations~(4), (1) and (3) of \citet{McGetal} as follows:
%
\begin{equation}
g_{\rm obs} =
\frac{g_{\rm bar}}{1 - \e^{-\displaystyle{\sqrt{g_{\rm bar}/g_\dagger}}}}
\label{eqn.g_dagger}
\end{equation}
with one fit parameter $g_\dagger$,
where $g_{\rm obs}$ is calculated from the observed rotation velocities by
%
\begin{equation}
g_{\rm obs} = \frac{v^2(R)}{R}~,
\label{eqn.obs}
\end{equation}
and $g_{\rm bar}$ from the gravitational
potential~$\Phi_{\rm bar}$ of the sum of the observed baryonic components:
%
\begin{equation}
g_{\rm bar} =
\left|\frac{\partial \Phi_{\rm bar}}{\partial R}\right|~.
\label{eqn.bar}
\end{equation}
The authors note in this context:
`There is no guarantee that $g_{\rm obs}$ should correlate with
$g_{\rm bar}$ when dark matter dominates.
Nevertheless, the radial acceleration relation persists for all galaxies
of all types.'

They find for the best fit parameter the acceleration\\
$g_\dagger = (1.20 \pm 0.02 \pm 0.24) \times 10^{-10}\m\s^{-2}$
with random and systematic uncertainties, respectively.
This acceleration~$g_\dagger$ agrees with
$a_0 \approx 1.2 \times 10^{-10}\m\s^{-2}$
given by \citet{Kroetal}, below which
the mass-discrepancy-acceleration correlation between $v_{\rm obs}$ and
$v_{\rm bar}$ deviates from the Newtonian value. For
accelerations~$a < a_0$, MOND would be applicable \citep{Mil83}. With this
assumption galactic DM is not required for a description of the dynamics of
galaxies \citep[cf.][]{FamMcG}.

As we will show in Sect.~\ref{s.multiple},
the flat rotational velocity curves of spiral galaxies can also be deduced
without DM contributions with the help of another model of gravitational
interactions based on a modified impact concept proposed for massive
bodies \citep{Wiletal}. The difficulties of the old impact theory
\citep[cf.][]{Fat90,Bop29} have been considered in the light of the Special
Theory of Relativity \citep[STR,][]{Ein05} and could be removed.
The basic idea is that impacting gravitons\,--\,originally called
quadrupoles\,--\,with no mass and a speed of light $c_0$ are absorbed by
massive particles and re-emitted with reduced energy~$T^-_{\rm G}$ according
to
%
\begin{equation}
T^-_{\rm G} = T_{\rm G}\,(1 - Y) ~,
\label{eqn.reduced}
\end{equation}
where $T_{\rm G}$ is the (very small) energy of a graviton in the background
flux and $Y$ ($0 < Y \ll 1$) is defined as the reduction
parameter. The corresponding momentum equation is
%
\begin{equation}
\vec{p}^-_{\rm G} = - \vec{p}_{\rm G}\,(1 - Y)~,
\label{eqn.reduced}
\end{equation}
with $|\vec{p}_{\rm G}| = T_{\rm G}/c_0$.
The equation implies that the diminished graviton is re-emitted in the
anti-parallel direction relative to the incoming one. This
geometry had
to be assumed in a study defining the mass equivalent
of the potential energy in a gravitationally bound two-body
system. An omni-directional emission, as assumed originally, led to conflicts
with energy and momentum conservation \citep{WilDwi15}.

Newton's law of gravitation could be explained with this model. Moreover,
a physical process causing the secular perihelion advances of the inner
planets and the Asteroid Icarus could be defined by assuming multiple
interactions within the Sun \citep{WilDwi}. The multiple interaction
process is obviously coupled with the presence of large mass conglomeration
and, therefore, galaxies might be the places to look for this process.
Despite the high mass values of spiral galaxies, they are
characterized by low acceleration levels in the outer regions due to their
large dimensions. A concentration of comparable amounts of matter would lead
to a more spherical geometry, which would not support the amplification process
proposed in Sect.~\ref{s.multiple}.
%
\begin{table*}[t]
\begin{tabular}{|c|c|c|c|} \hline
Physical quantity of galaxy & NGC\,7814 & NGC\,6503 & M\,33\\
\hline \hline
Mass, $M_{\rm bar}/\kg$ &
$1.0 \times 10^{41}$ & $2.1 \times 10^{40}$ & $\approx 1.8 \times 10^{40}$ \\
& \citet{CorSal} & \citet{McG05b} & \citet{Coretal}\\
& &  & $1.7 \times 10^{40}$ \\
&  &  & \citet{Kam17}\\
\hline
Velocity, $v_{\rm f}/(\km\s^{-1})$ &$\approx 210$  & 115 & 116 \\
& \citet{McGetal} & \citet{Greetal} & \citet{Coretal}\\
& (their Figure~2) & (their Figure~11) & (their Figure~15; MOND) \\
&&&\\
Mass, $M_{\rm bar}(v_{\rm f})/\kg$ &
$1.8 \times 10^{41}$ & $1.6 \times 10^{40}$ & $1.7 \times 10^{40}$\\
\hline
Best fit in Figs.\,1\,to\,3: &&&\\
$M^{\rm bf}_{\rm bar}/\kg$ &    $1.0 \times 10^{41}$
& $1.9 \times 10^{40}$  & $1.8 \times 10^{40}$  \\
\hline
\hline
\end{tabular}

\vspace{0.3cm}

Table~1:
Baryonic mass~$M_{\rm bar}$ and flat rotation velocity~$v_{\rm f}$
for the spiral galaxies NGC\,7814, NGC\,6503 and M\,33.
\end{table*}
%
\begin{table*}[t]
\begin{footnotesize}
\begin{tabular}{|c|c|c|c|c|c|c|c|} \hline
$Y$ & Left direction & Left side of disk &C& Right side of disk
& Right direction & Left or right sum & Factor~$f$ \\
\hline \hline
$1 \times$&&&&$z_0 = 100~\%~\longrightarrow$&$a~\Rightarrow$ &
$z \approx 0.37$&$f_1 =$\\
& $\Leftarrow~a$ & $\longleftarrow~100~\% = z_0$ &&& &
$a = {\rm e}^{\bar{x}/\lambda_0} = {\rm e}^{-1} = 0.37$& 0.37\\
\hline
$2 \times$  &  $\Leftarrow~(1 -a)\,a^2$ & $\longleftarrow~(1 - a)\,a$ &&
$\longleftarrow~(1 -a)$ & &$2 \times 0.085 \approx 0.17$ &$f_2 =$\\
&  &$(1 - a)~\longrightarrow$ && $(1 - a)\,a~\longrightarrow$ &
$(1 - a)\,a^2~\Rightarrow$&$2\,(1 - a)\,a^2$ & $0.54$\\
\hline
$3 \times$ & && & $(1 - a)^2\longrightarrow$ & $(1 - a)^2\,a~\Rightarrow$ &&\\
&  $\Leftarrow~~(1 - a)^2\,a$ & $\longleftarrow~(1 - a)^2 $ & && & &\\
& & $(1 - a)^2\,a~\longrightarrow$ && $(1 - a)^2\,a^2~\longrightarrow$ &
$(1 - a)^2\,a^3~\Rightarrow$&$3 \times 0.166 \approx 0.50$ &$f_3 =$\\
& $\Leftarrow~~(1 - a)^2\,a^3 $ &
$\longleftarrow~(1 - a)^2\,a^2$ && $\longleftarrow~(1 - a)^2\,a$ &
& 3\,$(1 - a)^2\,a\,(1 + a^2)$ & $1.04$\\
\hline
$4 \times$ &  $\Leftarrow~(1 - a)^3\,a^2$ &
$\longleftarrow~(1 - a)^3\,a$ && $\longleftarrow~(1 - a)^3$ & & &\\
& &  $(1 - a)^3 \longrightarrow$ &&
$(1 - a)^3\,a \longrightarrow$ & $(1 - a)^3\,a^2~\Rightarrow$ & &\\
&$\Leftarrow~(1 - a)^3\,a^2 $ &$\longleftarrow~(1 - a)^3\,a$ & && & &\\
& &  && $(1 - a)^3\,a \longrightarrow$ &
$(1 - a)^3\,a^2~\Rightarrow$ & &\\
&$\Leftarrow~(1 - a)^3\,a^4 $ & $\longleftarrow~(1 - a)^3\,a^3$ &&
$\longleftarrow~(1 - a)^3\,a^2$ & &$4 \times 0.073 \approx 0.29$ &$f_4 =$\\
&  & $(1 - a)^3\,a^2 \longrightarrow$ &&
$(1 - a)^3\,a^3\longrightarrow$ &
$(1 - a)^3\,a^4~\Rightarrow$ & $4\,(1 - a)^3\,a\,(2\,a + a^3)$
& $1.33$\\
\hline
$5 \times$  & & && $(1 - a)^4\longrightarrow$ & $(1 - a)^4\,a~\Rightarrow$
& &\\
& $\Leftarrow~~(1 - a)^4\,a$ & $\longleftarrow~(1 - a)^4 $ & & & &&\\
& & $2\,(1 - a)^4\,a~\longrightarrow$ && $2\,(1 - a)^4\,a^2~\longrightarrow$
&$2\,(1 - a)^4\,a^3~\Rightarrow$ & &\\
& & && $(1 - a)^4\,a^2~\longrightarrow$ &$(1 - a)^4\,a^3~\Rightarrow$& &\\
& $\Leftarrow~2\,(1 - a)^4\,a^3$ & $\longleftarrow~2\,(1 - a)^4\,a^2$ &&
$\longleftarrow~2\,(1 - a)^4\,a$ && &\\
& $\Leftarrow~(1 - a)^4\,a^3$ & $\longleftarrow~(1 - a)^4\,a^2$ &&& & &\\
&&  $(1 - a)^4\,a^3~\longrightarrow$ && $(1 - a)^4\,a^4~\longrightarrow$ &
$(1 - a)^4\,a^5~\Rightarrow$&$5 \times 0.084 \approx 0.42$& $f_5 = $ \\
& $\Leftarrow~(1 - a)^4\,a^5$ & $\longleftarrow~(1 - a)^4\,a^4$ &&
$\longleftarrow~(1 - a)^4\,a^3$ & &$5\,(1-a)^4\,a\,(1+3\,a^2+a^4)$
& {\bf 1.75} \\
\hline
$6\times$ & $\Leftarrow~(1 - a)^5\,a^2$&$\longleftarrow~(1 - a)^5\,a$&&
$\longleftarrow~(1 - a)^5$&&&\\
&&$(1 - a)^5~\longrightarrow$&& $(1 - a)^5\,a~\longrightarrow$&
$(1 - a)^5\,a^2~\Rightarrow$ &&\\
&$\Leftarrow~2\,(1 - a)^5\,a^2$& $\longleftarrow~2\,(1 - a)^5\,a$ &&&&&\\
&$\Leftarrow~3\,(1 - a)^5\,a^4$&$\longleftarrow~3\,(1 - a)^5\,a^3$&&
$\longleftarrow~3\,(1 - a)^5\,a^2$&&&\\
&&$3\,(1 - a)^5\,a^2\longrightarrow$ &&$3\,(1 - a)^5\,a^3\longrightarrow$&
$3\,(1 - a)^5\,a^4\Rightarrow$ &&\\
&&&& $2\,(1 - a)^5\,a~\longrightarrow$&$2\,(1 - a)^5\,a^2\Rightarrow$ &&\\
&$\Leftarrow~(1 - a)^5\,a^4$& $\longleftarrow~(1 - a)^5\,a^3$&&&&&\\
&$\Leftarrow~(1 - a)^5\,a^6~$&$\longleftarrow~(1 - a)^5\,a^5$&&
$\longleftarrow~(1 - a)^5\,a^4$&&&\\
&&$(1 - a)^5\,a^4~\longrightarrow$&&$(1 - a)^5\,a^5~\longrightarrow$&
$(1 - a)^5\,a^6~\Rightarrow$&$6 \times 0.049 \approx 0.29$&$f_6 = $\\
&&&&$~(1 - a)^5\,a^3~\longrightarrow$ &$(1 - a)^5\,a^4~\Rightarrow$&
$6\,(1-a)^5\,a^2\,(3+4\,a^2+a^4)$& $2.04$\\
\hline
\hline
\end{tabular}

\vspace{0.3cm}

Table~2: One-dimensional model of multiple interactions within a disk galaxy.
A large baryonic mass is assumed in the centre~C. Two opposite radial
directions to the left and right side are considered, where $z_0$
interactions lead to gravitons propagating with an energy loss of
$(1 \times Y)\,T_{\rm G}$. Several iterations inside the disk are
shown together with the gravitons leaving the galaxy in right
and left directions (marked by open arrows).
\end{footnotesize}
\end{table*}
\section{Details on the disk galaxies NGC\,7814, NGC\,6503 and M\,33} 
\label{Galaxies}
We will use three galaxies with different characteristics as examples. The
most relevant physical parameters of the galaxies for this study are compiled
in Table~1. The mass
data in the literature appear to be rather uncertain. Newer values
have been selected (when available). The calculations of the
mass~$M_{\rm bar}(v_{\rm f})$ have been performed with Eq.~(\ref{eqn.Tully_Fisher})
according to the BTF relation \citep{McG12}.
The best fit values of $M^{\rm bf}_{\rm bar}$ in the bottom row have been used in
Figs.\,\ref{fig.7814} to \ref{fig.M_33}.
Some additional details are given in the following subsections.

\subsection{NGC\,7814} 
\label{NGC_7814}
In the inner part, NGC\,7814 is a bulge dominated spiral galaxy
\citep[cf., e.g.,][]{Fraetal,Angetal}. The stellar and gas disks are seen
edge on. Its mass-to-luminosity ratio is very high. A substantial
DM contribution, closely coupled to the luminous component, seems, therefore,
necessary to explain the rotation curve in Table 3 of \citet{Fraetal}.
This rotation curve agrees with that of \citet{McGetal}. The
galactic radius is not well delimited and values between
$R_{\rm gal} = (2.8~{\rm to}~5.6) \times 10^{20}\m$ can be found in the
literature.

\subsection{NGC\,6503} 
\label{NGC_6503}
The late-type spiral galaxy NGC\,6503 is disk dominated and exhibits a regular
kinematical structure except for a remarkable drop of the stellar velocity
values in the central region \citep{BotGer,Greetal,McGetal}. The optical
radius is $R_{\rm gal} \approx 1.7 \times 10^{20}\m$
\citep{KoeBin}.

\subsection{M\,33} 
\label{M_33}
The spiral galaxy M\,33 has a rather complex structure
\citep[cf., e.g.,][]{Sei11}.
The total gas mass is of the same order as the stellar disk mass
according to \citet{Cor03}.
The optical radius is $R_{\rm gal} = 2.84 \times 10^{20}\m$
\citep{Kin15}.
\citet{Gir00b} finds that the results obtained for the
extended curve of M\,33 are not straightforward. At small
radii, there  is  a  wide  range  of  possible  models  with  or
without a disk dark component. The uncertainties related to the baryonic and
DM contributions in the central region are also highlighted by other authors
\citep[cf.][]{Cor07,Kametal15,Kam17}.

\section{Multiple interactions of gravitons in galactic disks} 
\label{s.multiple}
The large baryonic masses in galaxies will cause multiple interactions of
gravitons with matter if their propagation direction is within
the disk. For each interaction the energy loss of the gravitons is assumed to
be $Y\,T_{\rm G}$ \citep[for detail see Sect.\,2.3 of][]{Wiletal}.
The important point is that the multiple interactions occur only
in the galactic plane and not for inclined directions.
An interaction model is formulated in Table~2 indicating that an amplification
factor of approximately two can be achieved by six successive
interactions. The mean free path within
the plane of the disk be $\lambda_0$, then $z$ of $z_0$ gravitons will
not interact a second time within a distance~$x$ according to the equation:
%
\begin{equation}
z = z_0\,\e^{\displaystyle{-\frac{x}{\lambda_0}}}
\label{eqn.lambda}
\end{equation}
\citep[][p.\,160]{Wes56}. For $x = \lambda_0 = R_{\rm gal}$, we find with
$z_0 = 100$ that $z_0\,\exp(-1) = 100\,a = 37$ gravitons leave the
interaction region without a second reduction, whereas $1 - a$ will be
reflected with a loss of
$2 \times Y$. This process will be repeated several times and has been
formulated until a reduction of $6 \times Y$ is reached. The left column
gives the factors~$f_{\rm n}$ relative to a process with only
one interaction and $\lambda_0 = \infty$. As can be seen,
an amplification occurs for four or more interactions.
The process works, of course, along each diameter
of the disk and leads to a two-dimensional distribution of reduced gravitons.
%
\begin{figure}[t]
\begin{center}
\includegraphics[width=\columnwidth]{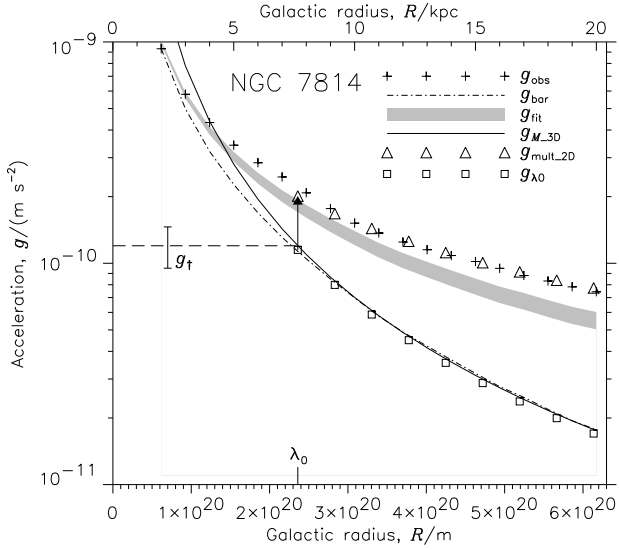}
\end{center}
\caption{Acceleration curves for NGC\,7814 versus galactic radius~$R$.
The data~$g_{\rm obs}$ and $g_{\rm bar}$ as well as the fit~$g_{\rm fit}$
with $g_\dagger = 1.2 \times 10^{-10}\m\s^{-2}$ are taken from \citet{McGetal}.
Outside of $\lambda_0$, corresponding to $g_\dagger$, the curves
$g_{M\_3{\rm D}}$ with $M^{\rm bf}_{\rm bar}$ of Table~1 (first column) and
$g_{\lambda_0}$ coincide with that of $g_{\rm bar}$, if a $1/R^2$ dependence
is assumed. The arrow indicates a multiplication factor~$f_5 = 1.75$.
Since the multiple interactions can only happen within the galactic plane,
the amplified acceleration decreases beyond~$\lambda_0$ proportional to $1/R$
in only two dimensions shown as $g_{\rm mult\_2D}$. It is in better
agreement with $g_{\rm obs}$ than $g_{\rm fit}$.}
\label{fig.7814}
\end{figure}
\subsection{Multiple interactions in NGC\,7814} 
\label{ss.mult_7814}
The bulge-dominated galaxy NGC\,7814 conforms best to the assumptions made
in Table~2. In Fig.~\ref{fig.7814}, the data presented by \citet{McGetal}
(first example in their Figure~2) are plotted as $g_{\rm obs}$ and
$g_{\rm bar}$, respectively, and compared with the fit of
Eq.\,(\ref{eqn.g_dagger}) as well as with the Keplerian acceleration in three
dimensions, $g_{M\_3{\rm D}}$, with the total baryonic mass~$M^{\rm bf}_{\rm bar}$
(see Table~1) at the centre. Below an acceleration of $g_\dagger$,
the curves for $g_{\rm bar}$ and $g_{M\_3{\rm D}}$ agree signaling that most
of the baryonic mass is indeed inside the corresponding radial distance
%
\begin{equation}
\lambda_0 = \displaystyle{\sqrt{\frac{G_{\rm N}\,M_{\rm bar}}{g_\dagger}}}~.
\label{eqn.lambda}
\end{equation}
In line with our assumptions in Table~2, we suggest that $\lambda_0$
provides an estimate of the mean free path
in the plane of the disk from the centre outwards, because
at this distance $g_{\rm obs}$ is larger than
$g_{\rm bar}$ by a factor of $f_5 = 1.75$ shown as arrow.\footnote{Note that
Eq.\,(\ref{eqn.g_dagger}) gives a ratio $g_{\rm obs}/g_{\rm bar} \approx 1.6$
for $g_{\rm bar} =  g_\dagger$.} Taking
into account the results of Table~2, five iterations of gravitational
interactions are thus required. Since the reflections can only occur within
the plane of the disk, the accelerations~$g_{\rm mult\_2D}$ at distances
greater than $\lambda_0$ decrease with $\lambda_0/R$ in two dimensions.
The agreement with $g_{\rm obs}$ is excellent.
\begin{figure}[t]
\begin{center}
\includegraphics[width=\columnwidth]{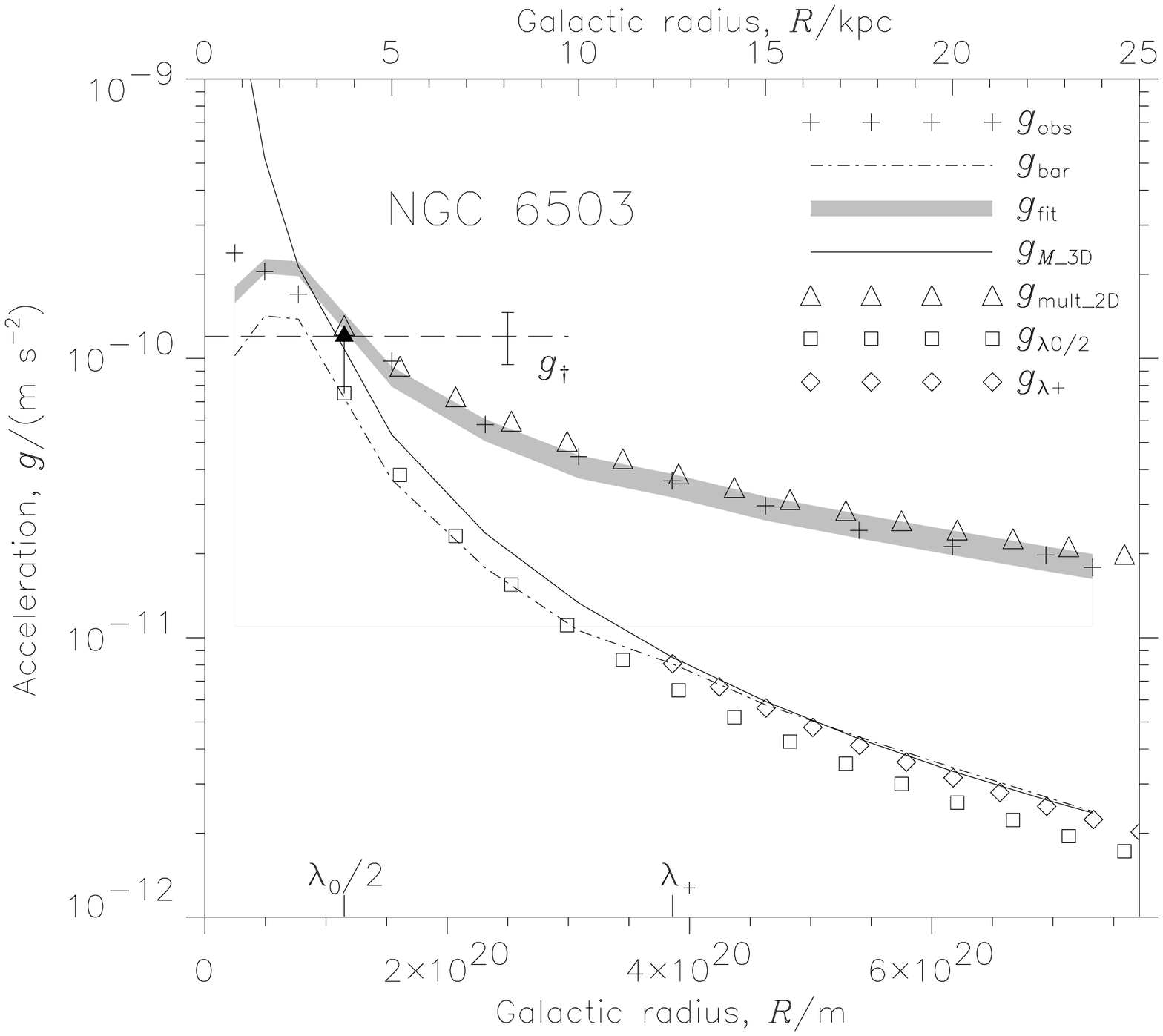}
\end{center}
\caption{Acceleration curves for NGC\,6503 versus galactic radius~$R$.
The data~$g_{\rm obs}$ and $g_{\rm bar}$ as well as the fit~$g_{\rm fit}$
with $g_\dagger = 1.2 \times 10^{-10}\m\s^{-2}$ are also taken from
\citet{McGetal}.
Between $\lambda_0$ and $\lambda_+ $, corresponding to
$g_{\rm bar} = 7.5 \times 10^{-11}\m\s^{-2}$ and
$\approx 10^{-11}\m\s^{-2}$, respectively, the curve
$g_{\rm bar}$ does not coincide with that of $g_{M\_3{\rm D}}$ for
$M^{\rm bf}_{\rm bar}$ of Table~1 (second column).
A substantial gas contribution between $\lambda_0/2$ and  $\lambda_+$
\citep[cf. Figure~2,][]{McGetal} is probably causing this behaviour. Only
beyond $\lambda_+$ the curves coincide, if a $\lambda_+/R^2$ dependence is
assumed. Irrespective of this complication,
a multiplication factor~$f_5 = 1.75$, indicated by the arrow at $\lambda_0/2$,
determines the amplified acceleration that decreases proportional to $1/R$
as $g_{\rm mult\_2D}$ in agreement with $g_{\rm obs}$ and $g_{\rm fit}$.}
\label{fig.6503}
\end{figure}

\subsection{Multiple interactions in NGC\,6503} 
\label{ss.mult_6503}
The disk-dominated galaxy NGC\,6503 also conforms quite well with the
assumptions
made in Table~2, although the central bulge is missing. Conceptually, this
could be modelled by omitting the column ``C'', i.e. the mean free path
$\lambda_0$ is related to the diameter of the galaxy and not to its radius.
In Fig.~\ref{fig.6503} the data presented by \citet{McGetal}
(second example in their Figure~2) are plotted as $g_{\rm obs}$ and
$g_{\rm bar}$, respectively, and compared with the fit of
Eq.\,(\ref{eqn.g_dagger}) as well as with the Keplerian acceleration in three
dimensions, $g_{M\_3{\rm D}}$, with the corresponding total baryonic
mass~$M^{\rm bf}_{\rm bar}$ (see Table~1) at the centre.
At an acceleration of $g_\dagger$, the curves for $g_{\rm obs}$,
$g_{M\_3{\rm D}}$ and $g_{\rm bar}$ multiplied by a factor of 1.75 agree at
a radial distance that we denote by~$\lambda_0/2$, taking the missing bulge
into account. This distance is
close to the galactic radius
$R_{\rm gal} = 1.7 \times 10^{20}\m$. Outside this radius~$g_{\rm mult\_2D}$
decreases with $(\lambda_0/2)/R$.
%
\begin{figure}[t]
\begin{center}
\includegraphics[width=\columnwidth]{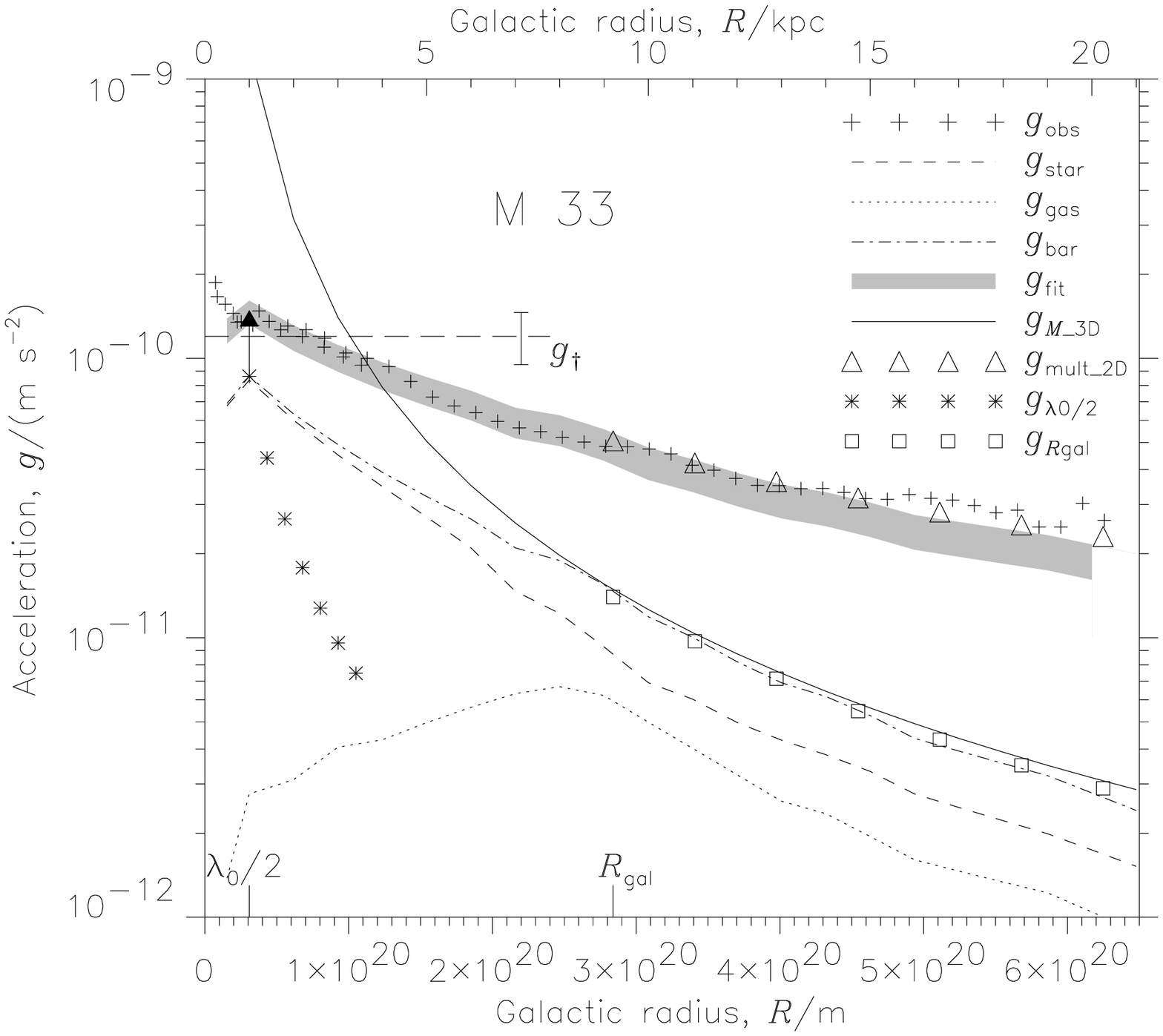}
\end{center}
\caption{Acceleration curves for M\,33 versus galactic radius~$R$.
The data for $g_{\rm obs}$ are from Table 1 of \citet{Coretal}
in agreement with Figure~6 of \citet{CorSal}, where the stellar
as well as gas contributions are taken from.
The fit~$g_{\rm fit}$, assuming $g_\dagger = 1.2 \times 10^{-10}\m\s^{-2}$, is
calculated with Eq.\,(\ref{eqn.g_dagger}) \citep[Eq.\,4 of][]{McGetal}.
A multiplication factor~$f_5 = 1.75$, indicated by the arrow at
$\lambda_0/2$, determines the amplified acceleration, however, the curves
$g_{M\_3{\rm D}}$ with $M^{\rm bf}_{\rm bar}$ of Table~1 (third column) and
$g_{R_{\rm gal}}$ coincide with that of $g_{\rm bar}$ only beyond
$R_{\rm gal} = 2.84 \times 10^{20}\m$.
A significant gas contribution \citep[cf. Figure~2,][]{McGetal} seems to be
responsible for this behaviour. The two-dimensional
fit~$g_{\rm mult\_2D}$ also agrees with $g_{\rm obs}$ only outside
the galactic radius.}
\label{fig.M_33}
\end{figure}
\subsection{Multiple interactions in M\,33} 
\label{ss.mult_33}
The galaxy M\,33 has a complex structure with a very large gas
contribution. Data from \citep{CorSal,Coretal} are shown in
Fig.~\ref{fig.M_33} and are again compared favourably with the fit of
Eq.\,(\ref{eqn.g_dagger}). As for NGC\,6503, there is no significant bulge
and thus the mean free path~$\lambda_0$ may again be related to the diameter
of the disk. Therefore, we plot $\lambda_0/2$ in this diagram
at a radial distance from the centre where
$g_{\rm obs}/_{\rm bar}$ is  $\approx 1.75$. Note, however,
that the complex structure of M\,33 casts some doubt on its position.
It could well be that $\lambda_0/2$ is much larger and an amplification
factor~$f_6 > 2$ or higher would be applicable.
Notwithstanding these difficulties
the curves~$g_{M\_3{\rm D}}$ and $g_{\rm bar}$ agree outside
of $R_{\rm gal}$, where $g_{\rm obs}$ can also be fitted very well
by~$g_{\rm mult\_2D}$.

\subsection{Common aspects of the galaxies NGC\,7814, NGC\,6503 and M\,33} 
\label{ss.common}
Both the bulge-dominated galaxy NGC\,7814 and the disk-dominated NGC\,6503
could be fitted quite well with our multiple interaction process and
a two-dimensional propagation of the affected gravitons. The mean free paths
of gravitons, $\lambda_0$, within the disks may be quite similar near
$\approx 2.4 \times 10^{20}\m$, if we take the different structures into
account. Even the mean free path in M\,33 might reach this value considering
the uncertainties involved here. In any case, the multiple interaction
process provides a good fit at distances greater $R_{\rm gal}$. Moreover, the
amplification factors of $f_n \approx 1.75~{\rm to}~2$ are appropriate at
acceleration levels of $g \approx 1 \times 10^{-10}\m\s^{-2}$ close to
$a_0$ \citep{Mil83,Mil15} and $g_\dagger$ \citep{McGetal}.

\section{Discussion and conclusion} 
\label{s.discuss}
The process of multiple interactions of gravitons with baryons in three spiral
galaxies leads to the observed flat velocity curves without the need to
invoke any DM. It thus provides a physical process for MOND
in two steps: (1) Multiple interactions of gravitons with baryons and
(2) a propagation confined to two dimensions.
Whether it can also be applied to other galaxies remains to be studied.

The process can also explain the BTF relation in
Eq.\,(\ref{eqn.Tully_Fisher}). Equating the centripetal acceleration
$g_{\rm obs}$ in Eq.\,(\ref{eqn.obs}) with the amplified baryonic
acceleration at $\lambda_0$ and taking into account its decrease with
$\lambda_0/R$, we find with Eq.\,(\ref{eqn.lambda})
%
\begin{equation}
g_{\rm obs}(R) = \frac{v^2(R)}{R} = \frac{f_5\,g_\dagger\,\lambda_0}{R}~.
\label{eqn.BTFR}
\end{equation}
For the flat rotation velocity~$v_{\rm f}$ it then follows
%
\begin{equation}
v^4_{\rm f} = f^2_5\,g_\dagger\,G_{\rm N}\,M_{\rm bar}~,
\label{eqn.BTFR}
\end{equation}
where $1/(f^2_5\,g_\dagger\,G_{\rm N}) = 4 \times 10^{19}\m^{-4}\s^4$.
This value is in reasonable agreement with the coefficient~$A$ in
Eq.\,(\ref{eqn.Tully_Fisher}) considering the discrepancies between
the mass values, in particular, for NGC\,7814.

It should be pointed out that the multiple interactions do not increase
the total reduction of graviton energy, because the total number of
interactions is determined by the (baryonic) mass of the gravitational centre
\citep[cf.][]{Wiletal}. A galaxy with enhanced gravitational acceleration
in two dimensions defined by the galactic plane, will, therefore,
have a reduced acceleration in directions inclined to this plane.
This might be relevant for the ``missing satellites problem'' mentioned in
Sect.~\ref{ss.S_gal}.

The ``core/cusp problem'' described in Sect.~\ref{ss.Core_cusp} is irrelevant
in the context of our discussion, because only baryonic matter is involved
in the multiple interaction process without any need for DM.

Finally, an effect of gravitational lensing should be touched upon, although
a detailed discussion is beyond the scope of this study.
It is, however, of interest that weak lensing signals
for prolate clusters can be twice that predicted under certain geometric
conditions \citep{Sch12}.

We want to conclude with a statement by \citet{Sotetal}
made in the context of gravitational theories in
`A no-progress report': `[...] it is not only the
mathematical formalism associated with a theory that is important, but the
theory must also include a set of rules to interpret physically the
mathematical laws'.


\acknowledgments
This research has made extensive use of the Astrophysics Data System (ADS).
Administrative support has been provided by the Max-Planck-Institute for
Solar System Research.

\end{document}